\documentclass[reprint,superscriptaddress,amsmath,amssymb,aps, prl,longbibliography ]{revtex4-2}

\usepackage{graphicx}
\usepackage{dcolumn}
\usepackage{bm}
\usepackage[colorlinks=true,allcolors=blue]{hyperref}
\usepackage{newtxtext,newtxmath}
\usepackage{microtype}
\usepackage{siunitx,xcolor,amsmath,amssymb}

\newcommand{\tc}{T_\mathrm{cook}}
\newcommand{\tr}{T_\mathrm{room}}

\newcommand{\partfigref}[2]{Fig.\,\hyperref[#1]{\ref*{#1}(#2)}}

\newcommand{\eg}{\textit{e.g.}}
\newcommand{\ecfp}{Edinburgh Complex Fluids Partnership and School of Physics and Astronomy, The University of Edinburgh,\\ James Clerk Maxwell Building, Peter Guthrie Tait Road, Edinburgh EH9 3FD, United Kingdom}
\newcommand{\eng}{School of Engineering, The University of Edinburgh, Alexander Graham Bell Building, Thomas Bayes Road, Edinburgh EH9 3FG, United Kingdom}

\newcommand{\titletext}{ Cooking crystalline candies and the ductile to brittle transition in concentrated suspensions}
\newcommand{\authors}{\author{Andreia F. Silva}
\email{andreia.silva@ed.ac.uk}
\author{James A. Richards}
\email{james.a.richards@ed.ac.uk}
\author{Fiona Jeffrey}
\affiliation{\ecfp}
\author{Rory E. O'Neill}
\affiliation{\ecfp}
\author{Daniel J. M. Hodgson}
\affiliation{\ecfp}
\author{Christopher Ness}
\affiliation{\eng}
\author{Wilson C. K. Poon}
\email{w.poon@ed.ac.uk}
\affiliation{\ecfp}
}
\makeatletter 
    
\renewcommand\onecolumngrid{
\do@columngrid{one}{\@ne}%
\def\set@footnotewidth{\onecolumngrid}
\def\footnoterule{\kern-6pt\hrule width 1.5in\kern6pt}%
}

\renewcommand\twocolumngrid{
        \def\footnoterule{
        \dimen@\skip\footins\divide\dimen@\thr@@
        \kern-\dimen@\hrule width.5in\kern\dimen@}
        \do@columngrid{mlt}{\tw@}
}%

\makeatother 

\begin{document}

\title{\titletext}%
\authors
\date{\today}

\begin{abstract}
The existence and origin of the ductile to brittle transition in non-Brownian suspensions and pastes is under-explored despite the ubiquity of such materials in practical applications. We demonstrate the phenomenon in candies of sugar crystals in a water-protein-fat matrix prepared by boiling a sugar-cream-butter mixture (known as `fudge' in some countries). As cooking time or final cooking temperature increases, we observe a transition from a fluid to a ductile solid, then to a brittle solid that abruptly fractures in compression. We propose that this is driven by rising solid sugar crystal volume fraction, and indeed find the same sequence of behaviour in a suspension of non-Brownian calcite particles as the solid fraction moves from frictional jamming to random close packing. Particle-based simulations reveal the  sensitivity of the observed phenomenon to boundary conditions. 
\end{abstract}
\maketitle

The ductile to brittle transition (DBT), triggered at low temperatures in many materials~\cite{Orowan1949}, has important practical implications. In the unusually cold North Atlantic of April 1912, the steel of the \textit{Titanic} fractured more brittly than modern ship-building steel; rapid sinking led to  catastrophic loss of life~\cite{Felkins1998,Deitz2012}. The DBT in the Earth's crust has seismological significance~\cite{Davarpanah2023}. And modern polymeric components are designed to place the DBT below operational temperatures~\cite{Ward2013}. 

There is increasing recent interest in the DBT in amorphous non-polymeric soft materials~\cite{Divoux2024}, \eg, in fibrous networks~\cite{Luo2018} and in printed 2D-lattice metamaterials~\cite{Berthier2019}. A key model here is the `jammed soft glass', where theory and simulation show that increasing annealing reduces stress inhomogeneity and transforms ductile to brittle yielding~\cite{ozawa2018random}, the latter being associated with shear localisation~\cite{barlow2020ductile}.

There has been little or no study of DBTs in particulate suspensions, which are ubiquitous in applications from concrete~\cite{richards2024fresh} to chocolate~\cite{Blanco2019}. Here, we study one such material, and find a DBT under compression in a confectionery consisting of a dispersion of sucrose crystallites in a complex matrix~\cite{Mcgee1992}. It is triggered by increasing final cooking temperature, $\tc$, which increases the crystal volume fraction $\phi$ as more water is boiled off. Comparison with a calcite powder paste~\cite{Richards2021} suggests that the DBT is generic in concentrated non-Brownian suspensions. Inspired by soft glasses, we hypothesise that this is driven by decreasing stress inhomogeneity as $\phi$ increases from frictional jamming to random close packing. Computer simulations show that how the DBT presents itself observationally is highly dependent on boundary conditions~\cite{Divoux2024}. 

\begin{figure}
    \includegraphics[height=3.5cm]{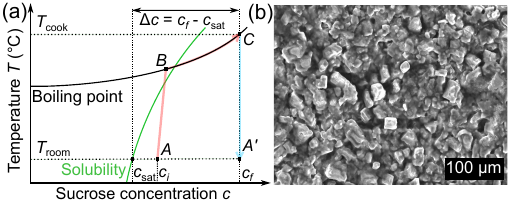}
    \caption{\label{fig:cook} Cooking fudge. (a)~Schematic thermal and compositional cycle, $ABCA^\prime$, from initial to final sucrose concentration, $c_i$ to $c_f$ and crystallisation. The black and green curves give the boiling point elevation, controlling $c_f$ at final temperature $\tc$, and solubility limit, which controls the saturation limit at $\tr$, $c_{\rm sat}$ (after real data~\cite{Sucrose}). The volume fraction of sucrose crystallites is controlled by $\Delta c = c_f - c_{\rm sat}$. (b)~Final microstructure of sucrose crystallites, in scanning electron micrograph of commercial fudge sample (JEOL JSM-6010PLUS/LV, gold coated sample at \SI{20}{\kilo\volt}) scale bar \SI{100}{\micro\metre}.} 
\end{figure}

Sugar-and-dairy confections are found globally, from Indian burfi~\cite{chetana2010effect} to Scottish tablet. Such `fudges' may show varied textures~\cite{Altan2021}. Our `minimal recipe' consists of sucrose (White Caster Sugar, Sainsbury), condensed milk (Nestle Carnation), golden syrup (N01D, Sucranna) and butter (Arla Pro), Table~\ref{tab:composition}. In a bespoke apparatus~\cite{Weir2016}, the system was heated at \SI{1.8}{\celsius\per\minute} towards \SI{100}{\celsius} then held until the fudge mixture reached \SI{90}{\celsius}. The mixture was then heated further to a series of temperatures between $\tc = \SIrange{108}{120}{\celsius}$, while stirring at 50\,rpm. Samples were cooled and beaten (i.e., vigorously agitated) to control nucleation~\cite{hartge2024rheo}, then transferred to silicone moulds to set for \SI{24}{\hour}.

In this process, \partfigref{fig:cook}{a}, a mixture with initial sucrose concentration $c_i$ at room temperature $T_{\rm room}$ is heated to boiling with little water loss ($AB$ near vertical). Next, substantial water loss ensues as the system traces the boiling point curve, $BC$, crossing the solubility boundary with little or no nucleation. Heating stops at $\tc$. The system at the final sucrose concentration $c_f$ cools along $CA^\prime$ back to $T_{\rm room} = \SI{20}{\celsius}$, where it is highly supersaturated and an amount of crystals $\propto \Delta c = c_f - c_{\rm sat}$ is precipitated. Using data on sugar solubility and density, we estimate $\phi(\tc)$ from $\Delta c$, Table~\ref{tab:volumefraction} (see End Matter for calculations). The final sucrose crystals are $\lesssim \SI{50}{\micro\meter}$, \partfigref{fig:cook}{b}.

\begin{table} 
    \caption{\label{tab:composition}Initial composition of `minimal fudge' recipe.}
    \begin{ruledtabular}
    \begin{tabular}{lcccccccc}
         & Sucrose & Lactose & Glucose & Protein & Fat  & Water & Other & Total \\
    Wt\,\% & 61.1 & 4.8 & 1 & 3.5 & 14 & 14.9 & 0.7 & 100 \\
    \end{tabular}
    \end{ruledtabular}
\end{table}

\begin{figure}
    \includegraphics[width=0.95\columnwidth]{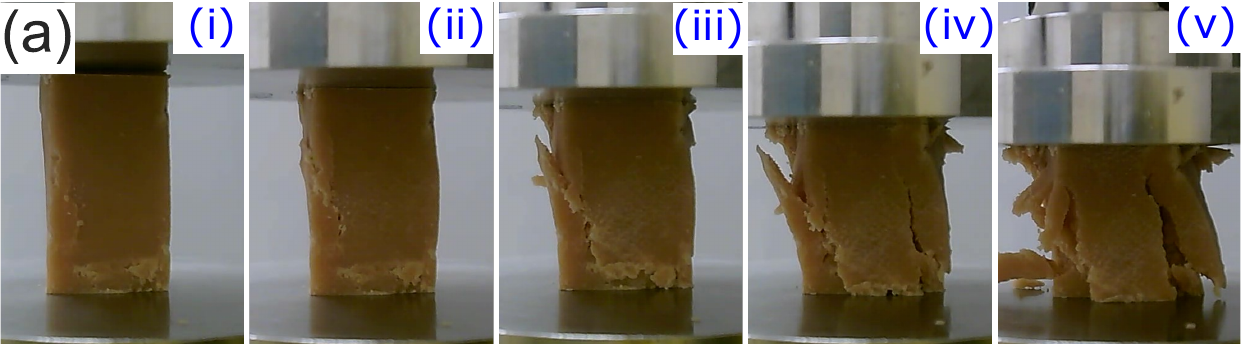}\\
    \includegraphics[width=0.95\columnwidth]{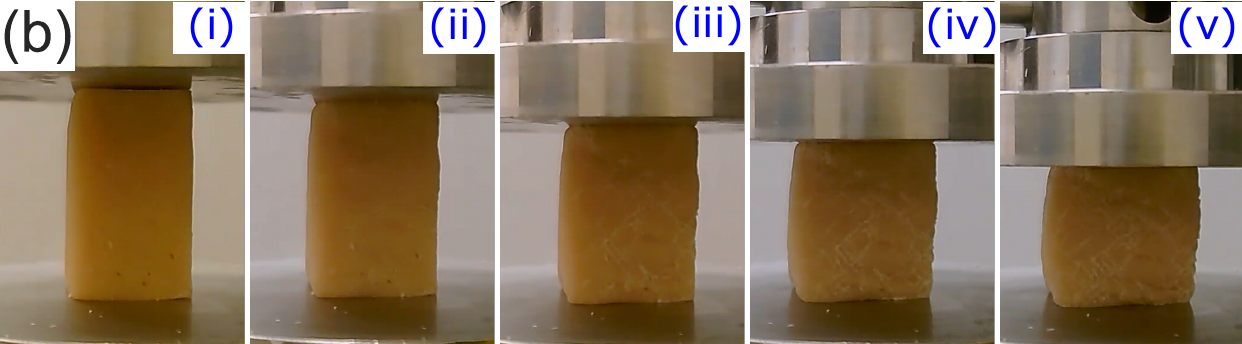}\\
    \includegraphics[width=1\columnwidth]{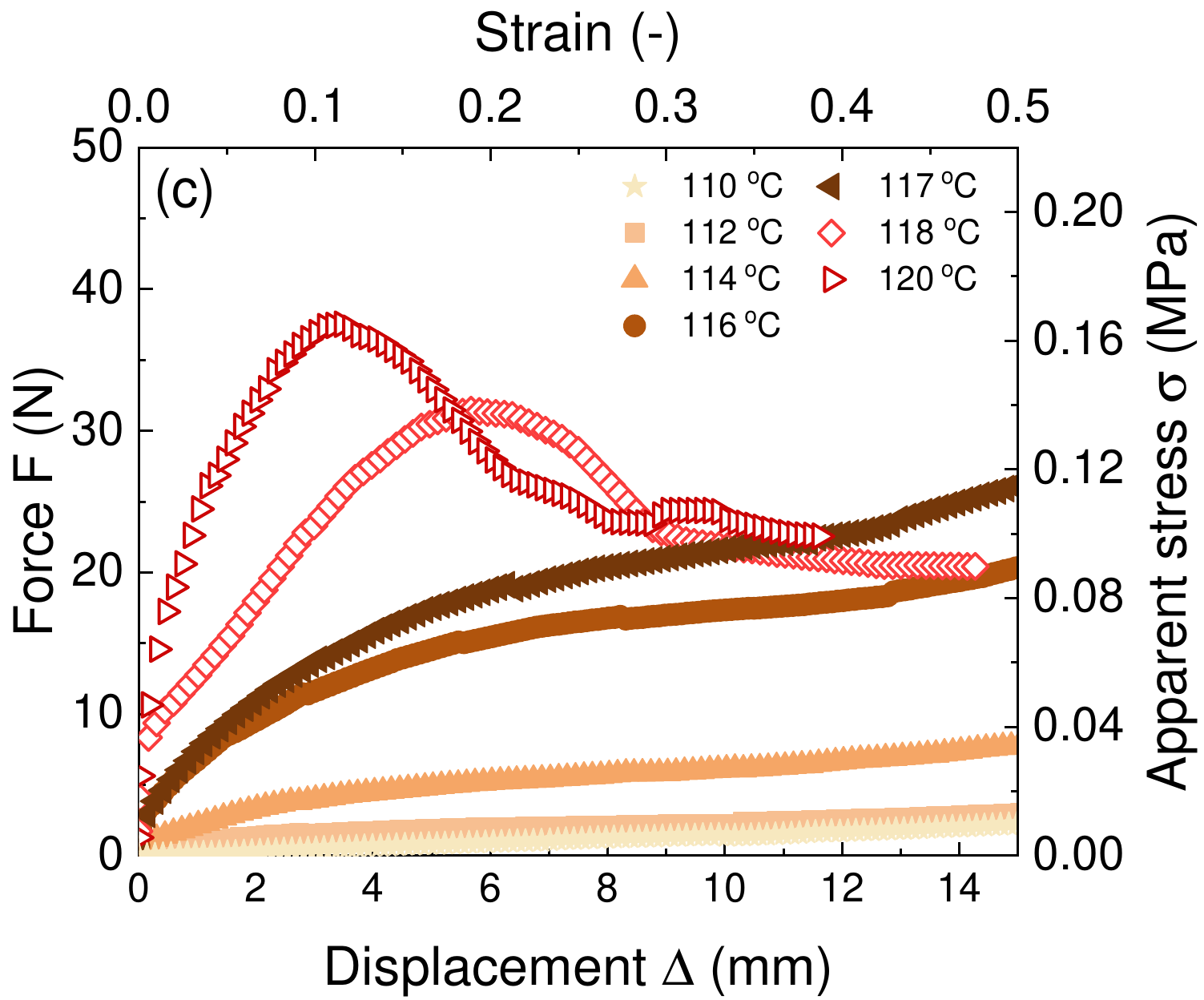}
    \includegraphics[width=0.49\columnwidth]{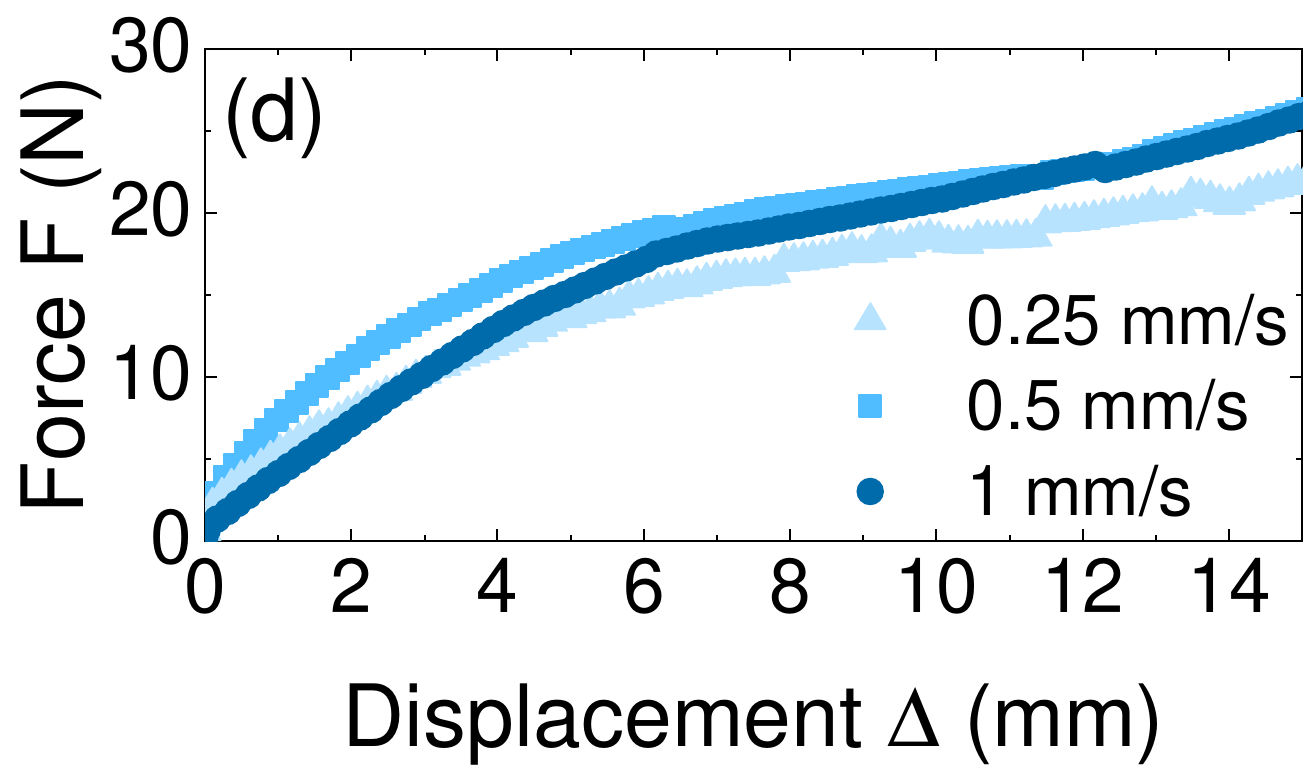}
    \includegraphics[width=0.49\columnwidth]{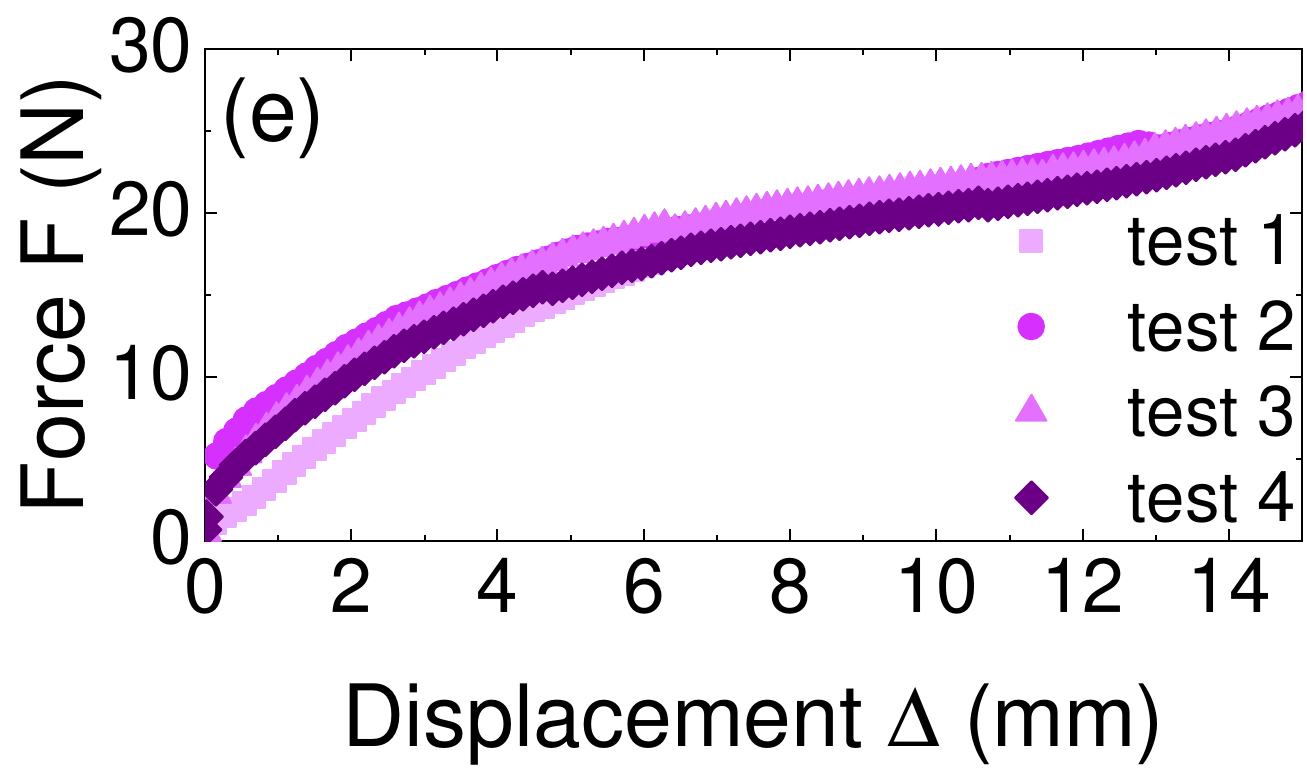}
    
    \caption{
    Compressional response of fudge cooked at increasing $\tc$. Images with (a)~$\tc  = \SI{120}{\celsius}$ and (b)~\SI{114}{\celsius} at increasing strain, (i)--(v). (c)~Force-displacement curves, $F(\Delta)$, and derived apparent stress-strain response, $\sigma(\epsilon)$ on top-right axes, for varying $\tc$, median curves presented from $\geq 3$ repetitions. (d)~Response with changing speed, 0.25, 0.5 and \SI{1}{\milli\metre\per\second} (light to dark symbols) for $\tc = \SI{117}{\celsius}$. (e)~Corresponding repetitions at $v = \SI{0.5}{\milli\metre\per\second}$.} 
    \label{fig:fudge}
\end{figure}

We compressed an $H = \SI{30}{\milli\meter}$ high sample with square cross-section (of side $L = \SI{15}{\milli\meter}$) between parallel plates in a universal testing machine (Lloyd/Ametek LS5, \SI{1}{\kilo\newton} load cell). After contact, the upper plate was lowered at speed $v =\SI{0.5}{\milli\meter\per\second}$ to mimic slow mastication. We recorded videos and force-displacement curves, $F(\Delta)$, and derive the apparent stress, $\sigma = F/L^2$, and strain, $\epsilon = -\Delta /H$.  A `well-cooked' sample [$\tc = \SI{120}{\celsius}$ \partfigref{fig:fudge}{a}], yields abruptly, first showing an $\approx \SI{45}{\degree}$ fracture from left to right, indicating failure along planes of maximum shear stress~\cite{Scholz2019}; later, a near-vertical crack rises from the base, typical of compressive brittle failure~\cite{zhou2018understanding}, and fragments fall off. An `undercooked sample' [$\tc = \SI{114}{\celsius}$ \partfigref{fig:fudge}{b}], compresses gradually, barrels, and displays multiple local fracture lines at $\approx \pm \SI{45}{\degree}$. In rocks, these features distinguish brittle and ductile fracture~\cite{Davarpanah2023}. 

Turning to the force-displacement data, $F(\Delta)$, \partfigref{fig:fudge}{c}, we find at the highest cooking temperature used, $\tc = \SI{120}{\celsius}$ ($\phi \approx  0.58$, open red triangles), there is a near-linear regime that peaks at $F \approx \SI{38}{\newton}~(\sigma \sim \SI{0.17}{\mega\pascal}$) followed by a drop. Such a `stress overshoot' is characteristic of brittle fracture. With decreasing cooking temperature, we observe an abrupt transition at $\tc = \SI{117}{\celsius}$ ($\phi \approx 0.56)$~[filled brown triangles]. Now, $F(\Delta)$ increases monotonically and then plateaus, before rising again as $\Delta \rightarrow \SI{15}{\milli\meter}$ due to lateral squeeze flow between the plates~\cite{engmann2005squeeze}. This ductile response is independent of $v$, \partfigref{fig:fudge}{d}, and is reproducible, \partfigref{fig:fudge}{e}. Decreasing $\tc$ from \SIrange{117}{112}{\celsius} and $\phi$ to 0.53, we find ductile solids of decreasing mechanical strength, \partfigref{fig:fudge}{c}~(dark to light brown). For $\tc < \SI{110}{\celsius}$ we reach the limit of compression testing set by sample weight, $\rho g H \sim \mathcal{O}(\SI{1}{\kilo\pascal})$.

\begin{figure}[tb]
    \centering
    \includegraphics[width=0.48\columnwidth]{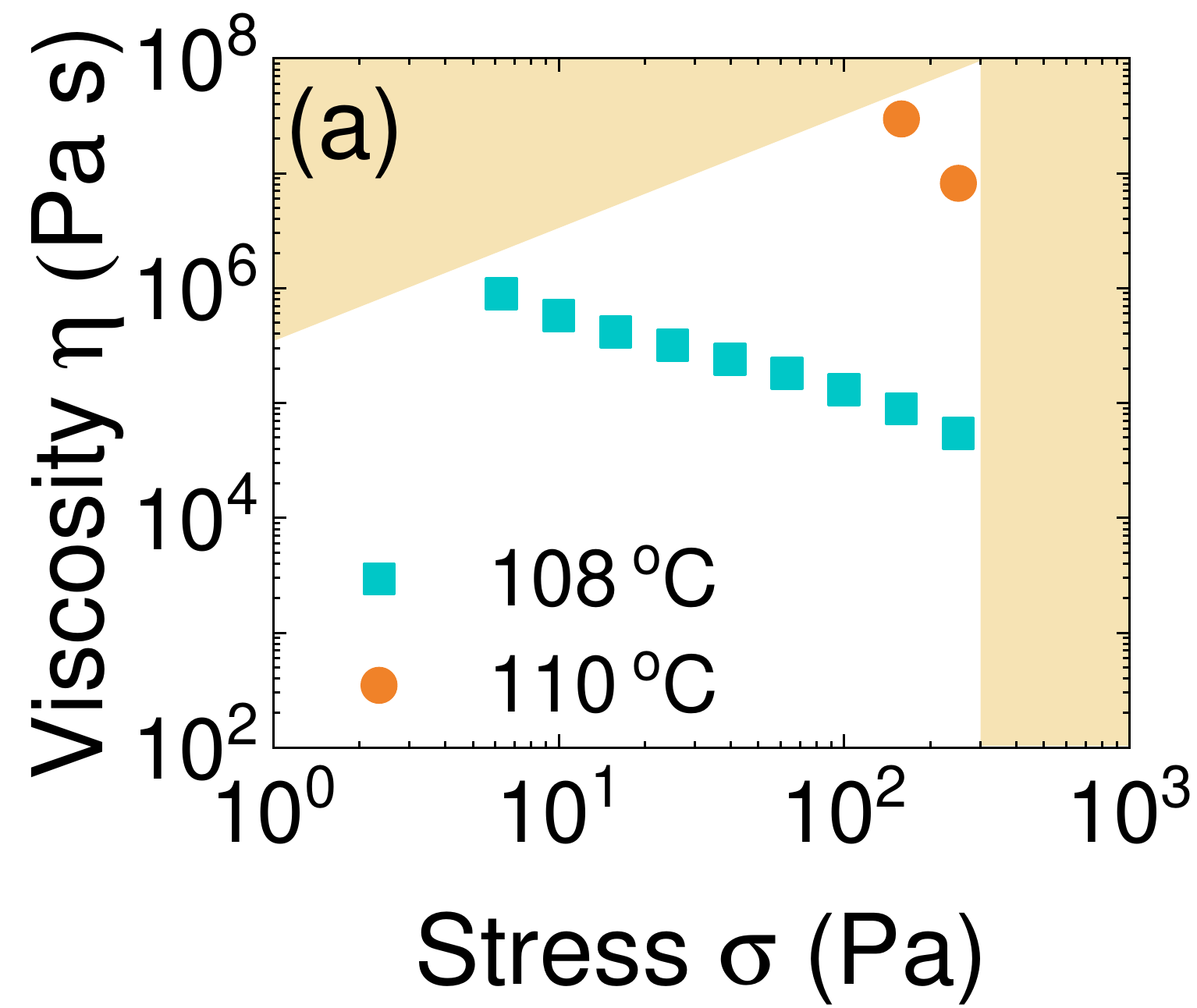}
    \includegraphics[width=0.48\columnwidth]{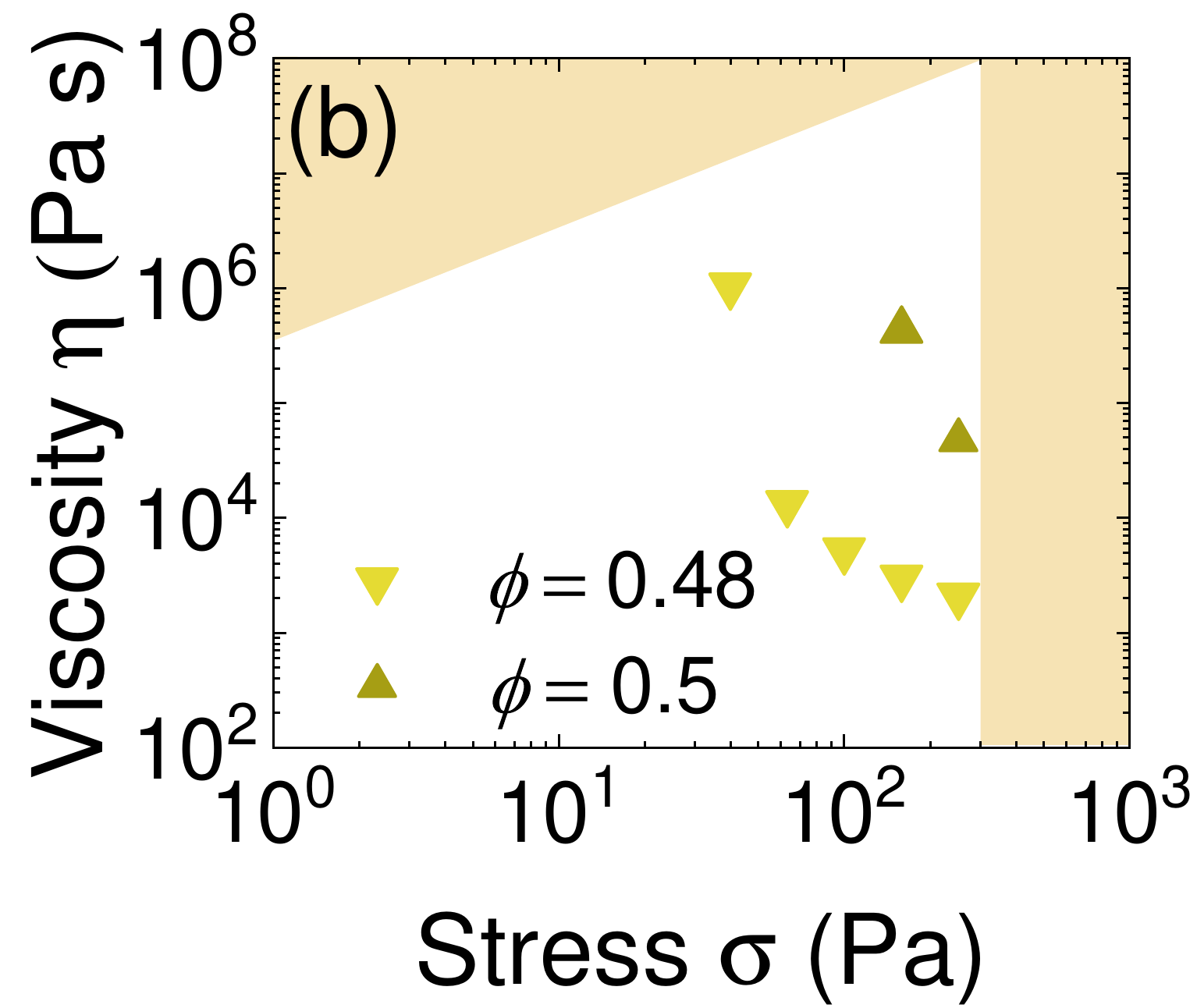}\\
    \caption{Shear rheology of fluid-solid transition. (a) Viscosity, $\eta$, with applied stress $\sigma$ for fudge at cooking temperatures $\tc=\SI{108}{\celsius}$ (blue squares) and \SI{110}{\celsius} (orange circles). Shading, measurement limits set by minimum resolvable strain rate (0.1\% over \SI{1000}{\second}) or sample fracture. (b)~Shear rheology for calcite suspensions at volume fractions $\phi^{c} =0.48 $ (light symbols) and 0.50 (dark).}
    \label{fig:rheo}
\end{figure}

We therefore next performed shear rheology [Kinexus Ultra+ (NETZSCH), 40\,mm cross-hatched parallel plates, \SIrange{2.0}{2.5}{\milli\meter} gap] on samples coated in mineral oil to prevent drying and sheared with a decreasing stress sweep from $\sigma = \SIrange{250}{1}{\pascal}$ at 5~pts/decade with \SI{1000}{\second} per step and averaging over the final \SI{100}{\second}. Higher $\sigma$ led to sample fracture~\cite{Blanco2019}. Decreasing $\sigma$ distinguishes between elasticity [negative strain of $\mathcal{O}(0.1\%)$] and viscous flow (positive strain), giving a measurable rheology `window', Fig.~\ref{fig:rheo} (shaded regions). 

\begin{table}[tb] 
    \caption{\label{tab:volumefraction} Volume fraction of crystals, $\phi$, for fudge cooked to different temperatures, $\tc$. Fat is neglected, assuming that it is dispersed at a large scale compared to the sucrose crystals, as in biscuit dough~\cite{pareyt2010impact}.}
    \begin{ruledtabular}
    \begin{tabular}{lcccccccc}
    $\tc$ (\si{\celsius})    & 108  & 110 & 112  & 114 & 116 & 117 & 118 & 120 \\
    $\phi$ & 0.51 & 0.52 & 0.53 & 0.54 & 0.55 & 0.56 & 0.57 & 0.58 \\
    \end{tabular}
    \end{ruledtabular}
\end{table}

Samples for $\tc > \SI{110}{\celsius}$ were too rigid to load homogeneously. While loading was possible for $\tc = \SI{110}{\celsius}$, only limited flow occurs at the limit of the measurable range, \partfigref{fig:rheo}{a}~(orange circles). The $\tc = \SI{108}{\celsius}$~(blue squares) sample is a shear-thinning fluid that flows across a wide range of stress, consistent with a non-Brownian suspension some way below the frictional jamming point, $\phi_\mu$~\cite{Guy2018, richards2020role, richards2024fresh}, and a transition from solid to liquid. At $\sigma = \SI{160}{\pascal}$, $\eta$ is $300 \times$ lower than the sample cooked at $\tc = \SI{110}{\celsius}$. 

Putting these results together, we conclude that our fudge undergoes two successive transitions with increasing $\tc$, or, equivalently, increasing $\phi$, Table~\ref{tab:volumefraction}. It transforms from fluid to solid at $\tc = \SI{110}{\celsius}$ and $\phi \approx 0.52$, and then from ductile to brittle solid at \SI{118}{\celsius} and $\phi \approx  0.57$. 

We hypothesise that increasing $\tc$ boils off more water and brings about the ductile to brittle transition by increasing the solid volume fraction. However, higher $\tc$ in our material has other effects such as altering the Maillard reactions between sugar and proteins~\cite{Hosry2025}. To test this hypothesis, we study suspensions of calcite crystalline powder (Eskal 500, KSL Staubtechnik, $\rho_p = \SI{2.7}{\gram\per\centi\meter^3}$) in an 80\,wt\% glycerol-water mixture ($\eta_f = \SI{0.061}{\pascal\second}$, $\rho_f = \SI{1.21}{\gram\per\centi\metre^3}$~\cite{glycerine1963physical}), where the calcite solid volume fraction, $\phi^c$, can be directly controlled from masses. This is a model suspension of hard, frictional particles with adhesive interactions~\cite{Richards2021}, for which critical volume fractions have been measured independently: the frictional jamming point $\phi^{c}_{\mu}= 0.48$~\cite{Gauthier2023} to 0.50~\cite{Richards2021} from the viscosity divergence in shear rheology and the maximal, random close packing $\phi^{c}_{\rm rcp}=0.60$ from dry powder compaction~\cite{Richards2021}. Shear rheology at $\phi^c = 0.48$ and 0.50, \partfigref{fig:rheo}{b}, confirms that a transition from shear-thinning across a wide $\sigma$ range to limited flow at the limit of measurement, evidencing $\phi^c_{\mu}$ and frictional jamming.

Calcite suspensions were prepared by repeated vortex and spatula mixing to a granule form~\cite{hodgson2022granulation}, allowing flow at $\phi^{c}>\phi^{c}_{\mu}$~\cite{garat2022using}, before compression into moulds. Compression tests on calcite pastes at $0.52 \leq \phi^{c} \leq 0.60$ followed a similar protocol to that for fudge. However, to accommodate sample shape variability, we used sample size measured after loading (initial $L = \SI{17.5}{\milli\metre}$, $H = \SI{35}{\milli\metre}$) and plate movement at $v = \SI{0.83}{\milli\metre\per\second}$ started above sample, defining $\epsilon = 0$ as the onset of measured stress. 

\begin{figure}[t]
    \includegraphics[width=0.8\columnwidth]{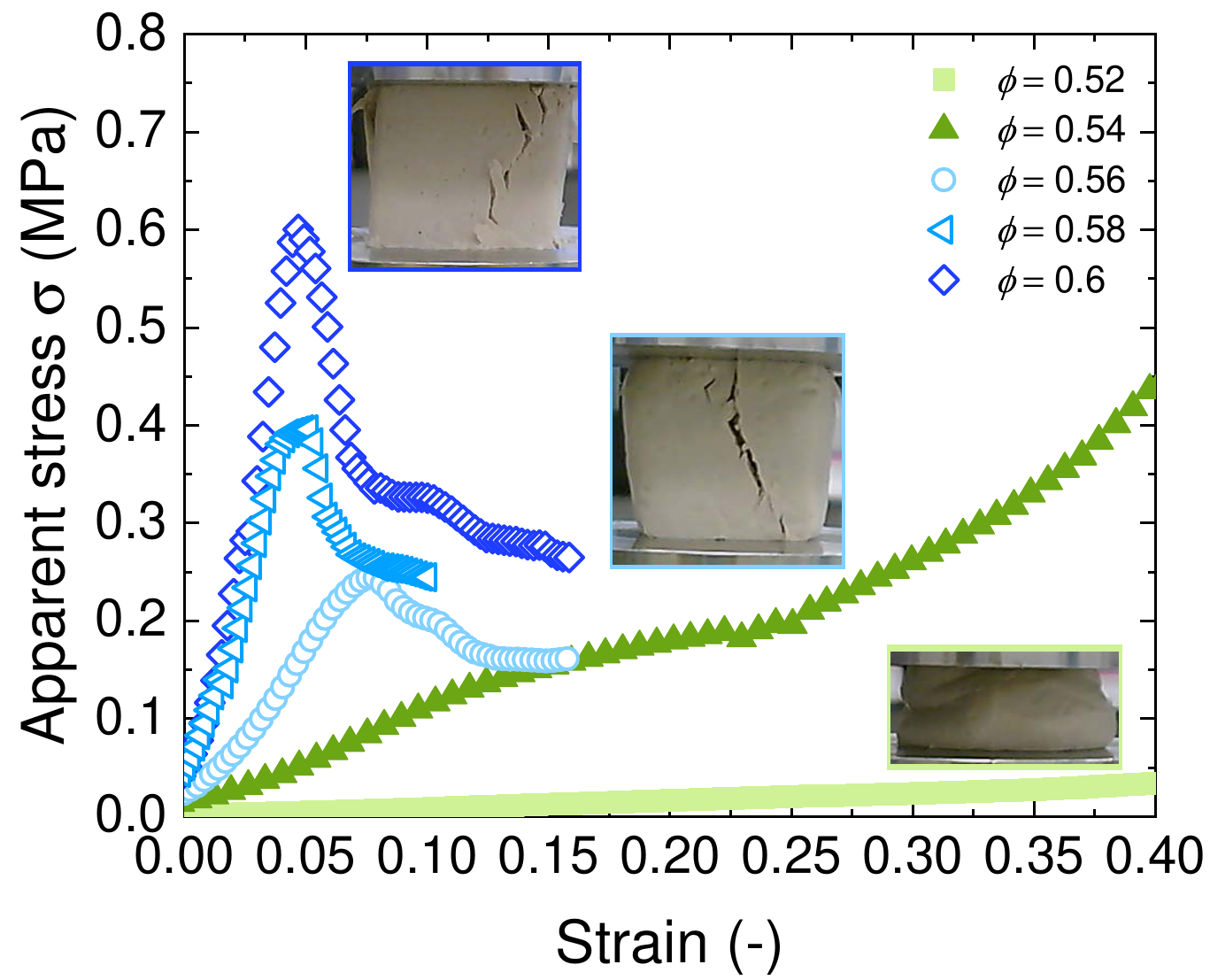}
    \caption{Calcite suspension compression tests. Apparent stress with engineering strain, $\sigma(\epsilon)$, for increasing $\phi^c$. Symbols, median curve for $\geq 3$ repeats with ductile yielding (filled, green) and brittle (open, blue). Inset: images of yielding at $\phi^c = 0.52$, 0.56 and 0.60.}
    \label{fig:calcite}
\end{figure}

At $\phi^{c} = 0.52$, the calcite paste shows ductile yielding, with a short yield stress plateau in $\sigma(\epsilon)$ at \SI{0.02}{\mega\pascal} at $\epsilon = 0.3$, before transitioning into squeeze flow with a rising $\sigma$, Fig.~\ref{fig:calcite}~(light green filled). At a higher $\phi^{c} = 0.54$ (dark green filled), the qualitative behaviour remains the same, but with the yield stress increasing to \SI{0.2}{\mega\pascal}; compare fudge from $\tc = \SIrange{112}{117}{\celsius}$, \partfigref{fig:fudge}{c}. At $\phi^{c} = 0.56$ the calcite paste enters the brittle transitional regime, with the appearance of a stress drop after a linear rise to \SI{0.25}{\mega\pascal} (light blue open). At greater $\phi^{c}$ up to 0.60 (dark blue open), the paste becomes increasingly stiff, with a steeper initial rise in stress up to \SI{0.6}{\mega\pascal}, and more brittle, with a larger stress drop. A corresponding onset of axial splitting is observed, Fig.~\ref{fig:calcite}\,(inset).

The close similarities between the observed phenomenology in fudge with increasing $\tc$, Fig.~\ref{fig:fudge}, and a calcite paste with increasing solid content $\phi^{c}$, Fig.~\ref{fig:calcite}, suggests that non-Brownian suspensions under compression show a generic sequence of transitions: a fluid-solid transition at frictional jamming followed by a DBT before random close packing. The close coincidence of the DBT, $\phi = 0.57$ for fudge and $\phi^{c} = 0.56$, is likely fortuitous, given uncertainties in estimating absolute volume fractions, especially for fudge. However, these broadly similar values may both reflect how particle angularity reduces $\phi_{\mu/\rm rcp}$ relative to spheres~\cite{azema2012nonlinear, azema2013packings}.

We have argued recently~\cite{Divoux2024} that the phenomenology of DBT is sensitive to geometry and boundary conditions, so that what is seen in uniaxial compression in a universal testing machine and in (say) cone-and-plate rheometry may differ considerably. We now explore such sensitivity computationally by simulating the Newtonian dynamics of spheres of density $\rho$ with radii $a$ and $1.4a$ using LAMMPS~\cite{ness2023simulating}. Details are given in Ref.~\cite{Cheal2018}. 

\begin{figure}
    \centering
    \includegraphics{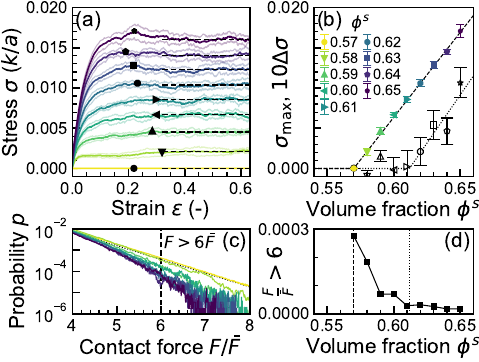}
    \caption{Frictional particle simulations. (a)~Strain dependence of deviatoric stress, $\sigma(\epsilon)$ with increasing volume fraction, $\phi^{s} = 0.57$ to 0.65 (light to dark). Solid lines, average of 10 realisations; shading, standard deviation; points, maximum stress $\sigma_{\max}$ for $\epsilon < 0.32 = \epsilon_{\max}/2$; dashed lines, $\bar\sigma$ for $\epsilon > \epsilon_{\max}/2$. (b)~$\phi$ dependence. Symbols: solid, $\sigma_{\max}$ $\propto \phi^s-0.570(3)$ for $\phi^s_{\mu}$ (dashed line); open, stress drop, $\Delta \sigma = \sigma_{\max}-\bar{\sigma} \propto \phi^s - 0.61(1)$, $10\Delta\sigma$ shown for clarity with error from standard deviation of $\bar\sigma$. (c) Normal contact force distribution, probability density ($p$) over $\epsilon_{\max}$ normalised by mean, $F/\bar{F}$. Dotted black line, experimental $p(F/\bar{F})$ for marginally jammed hard particles~\cite{mueth1998force}. (d)~Proportion of high force contacts with $F > 6\bar{F}$.}
    \label{fig:sim}
\end{figure}

Briefly, our periodic box contains 2000 particles, with 10 realisations for each volume fraction, $\phi^s$, with initialisation from a steadily sheared state. Particles $\alpha$ and $\beta$ experience a pairwise lubrication force with leading term $F^{\alpha,\beta}_i = \frac{\kappa}{h}\eta_f n_in_j(u^\beta_j-u^\alpha_j)$ with $\kappa$ a scalar function of the radii,
$n$ the centre-to-centre unit vector, $u^\alpha$ and $u^\beta$ the particle velocities and $h$ the surface-to-surface distance (indices denote Cartesian directions). The force is truncated when $h<10^{-3}a$, allowing direct particle contacts, which are Hookean with stiffness $k$ and frictional with coefficient $\mu_p=1$. We impose biaxial extension (uniaxial compression) with true rate $\dot{\gamma}$, setting $\rho\dot{\gamma}a^2/\eta=10^{-3}$ to minimise inertial effects. Particles experience drag forces that follow the affine deformation. The stress is computed by averaging the tensor product of pairwise vectors and forces. To compare with experiment we report $\sigma(\epsilon)$, the difference between compressional and extensional normal stresses as a function of uniaxial engineering strain,~\partfigref{fig:sim}{a}.

The stress response shows first a transition from fluid to ductile solid and the appearance of a significant stress response at $\phi^s_{\mu} = 0.57$ to 0.58, \partfigref{fig:sim}{a}~(yellow to green lines), as found in our experiments on both fudge and calcite suspensions when the solid fraction in each case increases beyond experimentally-estimated frictional jamming points. Thereafter, up to $\phi^s \leq 0.60$, $\sigma(\epsilon)$ has a steep, initially linear increase, followed by a monotonic plateau and ductile yielding for $\epsilon \gtrsim 0.1$ (light blue), again corresponding to experimental findings as solid content increases in either fudge and calcite suspensions. 

Beyond $\phi^s = 0.60$, the stress reaches a maximum as $\epsilon$ increases, $\sigma_{\max}$, \partfigref{fig:sim}{a} (filled symbols), before slightly dropping down to a plateau level, (dashed lines). The overshoot, $\Delta \sigma$, grows until $\phi_{\rm rcp}^s = 0.65$, \partfigref{fig:sim}{b}~(open symbols). The shape of the simulated overshoot differs from experiments, where, instead of a post-overshoot plateau, we observe a rapid and large post-overshoot drop in the stress. This difference between experiments and simulations is traceable to differing boundary conditions. The
abrupt decrease in stress after the overshoot observed in experiments is the consequence of macroscopic \SI{45}{\degree} and vertical fractures, which are not possible under the \emph{constant volume} periodic boundary conditions of our simulations; without macroscopic fracture, simulations therefore do \emph{not} show a large post-overshoot drop in stress, which experimentalists tend to take as a `signature' of brittle fracture. But this visually striking difference in the data, compare Fig.~\ref{fig:calcite} and \partfigref{fig:sim}{a}, does \emph{not} evidence any fundamental difference in physics, but only non-essential difference in boundary conditions.  


These differences notwithstanding, simulations do provide support for the role of decreasing stress inhomogeneities in triggering the DBT as $\phi$ increases from $\phi_\mu$ to $\phi_{\rm rcp}$, following the model of annealed glasses~\cite{ozawa2018random}. The force chain network for frictional systems, which is sparse and localised when marginally jammed by shear at $\phi \gtrsim \phi_{\mu}$~\cite{cates1998jamming}, becomes more uniform when approaching isotropic jamming~\cite{majmudar2005contact}, as occurs at $\phi_{\rm rcp}$. The distribution of normal forces in our discrete element method (DEM) simulations is consistent with this picture and matches the phenomenology of 2D sheared packings~\cite{behringer2019physics}: where $\Delta\sigma$ arises, the high-force tail becomes truncated as $\phi^s$ increases above $\phi^s_\mu$ up to $\phi^s=0.61$, \partfigref{fig:sim}{c}. The proportion of contacts with a large load compared to the mean ($F/\bar{F}>6$) reduces from $3\times10^{-4}$ to $\leq 3\times10^{-5}$, \partfigref{fig:sim}{d}.


Finally, we seek to understand the experimentally-observed peak stresses of $\lesssim \SI{1}{\mega\pascal}$, Fig.~\ref{fig:calcite}, using the magnitude of the peak stress found in simulations just before boundary conditions become critical, \partfigref{fig:sim}{a}, $\sigma_{\max} = 0.008k/a$ at $\phi^s=0.61$ rising to $0.016 k/a$ at $\phi^s_{\rm rcp}$. As particles interact via linear springs in these simulations, these peak stresses correspond to compressive strains due to particle overlap of $\delta \sim 0.01$. Calcite has Young's modulus $E \approx \SI{55}{\giga\pascal}$ and compressive strength $\sigma_c \approx \SI{200}{\mega\pascal}$, so that an upper bound for its compressive yield strain is $\sigma_c/E \lesssim 4 \times 10^{-3}$~\cite{Carmichael1982}. In practice, particle curvature concentrates stress, further reducing the yield overlap (for spheres) to $\delta_y \sim (\sigma_c/E)^2 \approx 10^{-5}\ll \delta$ ~\cite{rathbone2015accurate}. Our simulations therefore show that the particles in our experiments must be deforming plastically at the peak stress. This deduction is consistent with the observation that our fudge samples did not elastically recover even when deformed strictly within the linear regime ($\epsilon \ll 0.1$).  


In the plastic regime, it has been argued that inter-particle forces are linear with effective stiffness $k/a \sim \sigma_c$~\cite{thornton1997coefficient}. 
So, the peak stresses found in simulations, \partfigref{fig:sim}{a}, $\sigma_{\max} = 0.008k/a$ to $0.016 k/a$ at $\phi^s_{\rm rcp}$, convert to experimental stresses in our calcite suspensions of \SIrange{1.6}{3.2}{\mega\pascal}. These are higher than, but within an order of magnitude of, the experimentally-observed peak stresses of \SIrange{0.25}{0.6}{\mega\pascal}, Fig.~\ref{fig:calcite}, supporting the physical picture we have given of plastic contact deformation. The residual discrepancy may be due to the use of $\sigma_c$ for a high-purity calcite, whereas our material may be in the lower range of $\sigma_c$ observed~\cite{Carmichael1982}. 

To summarise, we have studied experimentally the compressive behaviour of two adhesive non-Brownian suspensions, `fudge', a confectionery of sugar crystals suspended in a soft matrix of proteins and fats~\cite{Mcgee1992}, and calcite suspended in glycerol-water~\cite{Richards2021}. As the solid volume fraction increases, each shows a fluid-solid transition associated frictional jamming followed by a clear ductile-to-brittle transition (DBT) displaying a `signature' abrupt stress drop upon yielding~\cite{Divoux2024} and yielding along a \SI{45}{\degree} axis followed by axial splitting and fracture~\cite{Davarpanah2023}. By directly varying the suspension concentration in our calcite suspension, we find jamming at $\phi_{\mu}^c = 0.52$,  and $\phi_{\rm DBT}^c = 0.56 < \phi_{\rm rcp}^c = 0.60$. In fudge, where increasing sugar crystal volume fraction was achieved by longer cooking time to boil off more water, we inferred the same sequence of transitions by backing out the solid loading indirectly. Particle-based simulations show a change in response at a critical point between frictional jamming and random close packing. However, the local rheology shows only a weak stress overshoot. This confirms the  suggestion~\cite{Divoux2024} that how DBT is manifested is highly sensitive to geometry and boundary conditions.



Analogy with the role of annealing in the DBT in soft glasses suggests the DBT in non-Brownian suspensions is associated with increasingly homogeneous stress distribution as $\phi$ increases. Changes in the particle contact force distribution support this picture, although probes of collective microscopic yielding events and stress propagation in jammed frictional materials are needed to confirm this. Our observation of differences between experiments and simulations of compressive yielding contributes to a growing body of evidence highlighting the key role of boundary conditions, e.g., the observation of shear thinning in simulations under fixed pressure but shear thickening at constant volume~\cite{ness2025nonmonotonic}, and of brittle tensile failure~\cite{Smith2010,Smith2015} but ductile shear yielding~\cite{barik2024role} in experiments.

For practical applications, our finding that the magnitude of the stress peak in the brittle fracture of granular suspensions is set by the compressive plastic strength of the particles brings the predictive design of the mechanical response of such materials one step closer. Separately, the peak stress at the DBT for fudge, $\sigma \sim \SI{0.1}{\mega\pascal}$, \partfigref{fig:fudge}{c}, is in the range applied in the buccal cavity to soft confectioneries~\cite{Ono2009}. Our findings should therefore contribute to understanding the mouthfeel of such confectioneries~\cite{civille2011food}. While the correlation between viscosity and mouthfeel has been widely studied~\cite{deblais2021predicting}, the role of complex yielding behaviour remains largely unexplored. Our study provides a foundation for future work in this direction.



\begin{acknowledgments}
An Interface Innovation Voucher funded work with Ochil Fudge Ltd., whose staff introduced us to fudge making. AFS and JAR were partly funded by the UK Engineering and Physical Sciences Research Council Impact Acceleration Account [EP/R511687/1 and EP/X525698/1]. JAR and WCKP thank participants at the 2023 Lorentz Center workshop on Yield Stress and Fluidization in Brittle and Ductile Amorphous Systems~\cite{Divoux2024} for discussions. 
\end{acknowledgments}

\bibliography{DBT}

\onecolumngrid

\begin{center}
    \textbf{\large End Matter}
\end{center}

\twocolumngrid

\paragraph{Fudge composition}The final fudge is assumed to be a protein network/dispersion in a sugar suspension in which the crystals are mainly sucrose with a minority lactose component, and the `solvent' is a saturated sucrose solution with fat spread over a larger scale. We want to estimate the volume fraction of sugar crystals in the solvent. The values of sucrose- and lactose physical properties are from \citet{Sucrose} unless otherwise stated. We assume a final temperature of \SI{20}{\celsius}.

As sucrose is more soluble in water than lactose \cite{Hartel2017}, to first approximation we will consider only sucrose. Consider \SI{100}{\gram} of a sucrose-water solution that is boiling at \SI{120}{\celsius}. It contains \SI{86.9}{\gram} of sucrose and \SI{13.1}{\gram} of water. At $\tr = \SI{20}{\celsius}$, it is possible to dissolve 
\SI{26.2}{\gram} of sucrose in our \SI{13.1}{\gram} of water, giving \SI{39.3}{\gram} of saturated solution. The remaining \SI{60.7}{\gram} of sucrose will be in the form of crystallites. The densities of saturated sucrose solution {(\SI{1.327}{\gram\per\cubic\centi\meter} \cite{Emmerich1994}) and sucrose crystals (\SI{1.588}{\gram\per\cubic\centi\meter}) give $ \SI{38.3}{\cubic\centi\meter}$ of crystallites in $\SI{29.6}{\cubic\centi\meter}$ of saturated sucrose solution. The volume fraction of crystals is therefore $\phi \approx \frac{38.3}{38.3+29.6} = 0.56$.

Next we consider other sugars. Proportionate to the recipe and composition in Table~\ref{tab:composition}, there is \SI{5.9}{\gram} of lactose and \SI{1.2}{\gram} of syrup-derived `inverted sugar' ($\sim$ 1:1 number ratio of glucose and fructose) mixed in with the sucrose-water mixture considered above. The inverted sugar acts to retard sucrose nucleation, but will likely remain solubilised. We will neglect its contribution. Assuming that a saturated solution of sucrose cannot dissolve any of the lactose, this leaves \SI{5.9}{\gram} of lactose crystals. The density of lactose crystals (\SI{1.546}{\gram\per\cubic\centi\meter}~\cite{Kaialy2011}) adds $\SI{3.80}{\cubic\centi\meter}$ of crystals and marginally increases our estimate to $\phi(\tc) = \SI{120}{\celsius} \approx \frac{38.3 + 3.80}{38.3 + 3.80 + 29.6} = 0.58$. Repeating such analysis for our fudge cooked to the other temperatures gives the $\phi(T_{\rm cook})$ values presented in Table~\ref{tab:volumefraction}. If we considered that a saturated sucrose solution dissolved lactose, our $\phi$ estimation would be up to 0.006 lower for all $T_{\rm cook}$ studied.
\end{document}